\documentclass[12pt]{iopart}

\usepackage{graphicx}                 
\usepackage{epstopdf}
\usepackage{color}                    
\usepackage[colorlinks=true,dvipdfmx]{hyperref}                 
\usepackage{makeidx}
\usepackage{fancyvrb}
\usepackage{bm}
\usepackage{color}

\begin{document}

\title[]{Implementation of LDA+DMFT with  pseudo-potential-plane-wave method}

\author{Jian-Zhou Zhao$^{1,2}$, Jia-Ning Zhuang$^1$, Xiao-Yu Deng$^1$, Yan Bi$^2$, Ling-Cang Cai$^2$, Zhong Fang$^1$, and Xi Dai$^1$}

\address{$^1$Institute of Physics, Chinese Academy of Sciences, Beijing 100190, People's Republic of China}+
\address{$^2$National key Laboratory of Shock Wave and Detonation Physics, Institute of Fluid Physics, China Academy of Engineering Physics, Mianyang, 621900, China}
\eads{daix@aphy.iphy.ac.cn}

\date{\today}

\begin{abstract}
In this paper, we propose an efficient implementation of combining Dynamical Mean field theory (DMFT) with electronic structure calculation based
on the local density approximation (LDA). The pseudo-potential-plane-wave method is used in the LDA part, which makes it possible to be applied
to large systems. The full loop self consistency of the charge density has been reached in our implementation which allows us to compute the 
total energy related properties. The procedure  of LDA+DMFT is introduced in detail with a complete flow chart. We have also applied our
code to study the electronic structure of several typical strong correlated materials, including Cerium, Americium and NiO. Our results fit quite well with
both the experimental data and previous studies.
\end{abstract}



\maketitle

\section{Introduction}\label{intro}
The first principle calculation based on the  density functional theory (DFT) with the local density approximation (LDA) and its generalization 
generalized gradient approximation (GGA) 
is very successful in predicting ground-state properties and band structures of a wide range of real materials.
However it is known for a long time that it is not sufficient to calculate the electronic structure of those strongly correlated materials
 (i.e. transition metal oxides, actinide and lanthanide-based materials, high $T_c$ superconductors)  by these methods alone even on 
the qualitative level.
In order to overcome these shortcomings of the traditional DFT-LDA scheme, remedies such as LDA+$U$ have been proposed, 
which can describe some of the strong correlated materials with long range order in their ground states. While even
LDA+$U$ can not be applied to many of the strongly correlated materials, for example, the paramagnetic Mott 
insulator phase and the materials containing unfilled f-shell with strong multiplet effects. 

Therefore it is important to have a first principle method which can be applied to the 
strongly correlated materials with the ability
to capture the dynamical properties, finite temperature properties and multiplet effects.
In the last two decades, the dynamical mean field theory (DMFT) has been developed very fast to be the standard tool to deal with the 
on-site correlation effect in the limit of large dimension \cite{Georges1996}. After being successfully applied to many model systems, DMFT
has been considered as a powerful tool to capture the on-site correlation effects based on the Hubbard like Hamiltonians containing both the local interaction terms and the  single particle Hamiltonians extracted from LDA. 
Therefore the LDA+DMFT method which combines the DFT-based band structure techniques with DMFT has been proposed and developed quickly in 
the last decade.  By DMFT the local correlation effect can be well described by the 
self energy, which has frequency dependence and in general takes the matrix form within the subspace spanned by the correlation orbitals.   Using
the Green's function containing self energy, many of the physical properties of strongly correlated materials can be calculated, i.e. the electronic 
spectral function, the total energy, the optical conductivity and the local spin susceptibility. 

Most of the LDA+DMFT calculation till now  have been performed by the partially self-consistent scheme, where the local self-energy is obtained by the DMFT calculation with a fixed LDA Hamiltonian generated from the fixed LDA charge density. Therefore, in this simplified scheme one neglects the inference of the strong on-site Coulomb interaction on the charge density. The above mentioned ``one-shot''  DMFT calculation works quite well
for the electronic structure. While in order to obtain reliable total energy related properties, i.e. the equilibrium  volume, the elastic constants and the
phonon frequencies, the LDA+DMFT scheme with full charge density self consistency is needed.
 Up to date,  there have been several fully self-consistent LDA+DMFT schemes \cite{Lechermann2006,Anisimov2007a} as well as actual implementations \cite{Savrasov2004,Minar2005,Pourovskii2007,Haule2010}.  As mentioned in \cite{Haule2010}, there are three major issues that have to be addressed in DFT+DMFT implementations: i) quality of the basis set, ii) quality of the impurity solvers, iii) choice of correlated orbitals onto which the full Green's function is projected. In the implementations of LDA+DMFT  mentioned above with full charge self-consistency , the basis set of the linear muffin-tin orbitals (LMTO) are used, and the local correlated orbitals are naturally chosen based on the muffin-tin orbitals . While on the other hand, there are very few implementations of  LDA+DMFT with full charge self-consistency are based on the plane wave methods.

In the present paper, we implement a full self-consistent LDA+DMFT scheme using pseudo-potential-plane-wave as the basis set, and we use either atomic orbitals or Wannier orbitals depending on different physics systems. The Hubbard-I approximation is used as the impurity solver in the present paper, but it can be replace by more precise solver like continuous time quantum Monte Carlo (CTQMC) for metallic systems. The present paper is organized as follows. In \sref{projection}, we describe the way to choose and construct the correlated orbitals onto which the local interactions exert. In \sref{ldadmft}, we show the full-loop flow and LDA+DMFT(Hubbard-I) formalism in detail. We apply this LDA+DMFT approach to several correlated materials such as Ce, Am, and NiO in \sref{benchmark}. Finally we conclude our work in \sref{conclusions}.

\section{Projection onto localized orbitals}\label{projection}
In the LDA+DMFT calculation, all the energy bands are divided into two groups.
And only the on-site interactions among those local orbitals are treated more precisely
by DMFT. Therefore the choice of localized orbitals
does affect the results obtained by LDA+DMFT method and becomes one of the 
important issues in LDA+DMFT.
A natural choice for the
local orbitals is a set of atomic like wave functions with
corresponding $d$ or $f$ characters. Typical atomic like local
orbitals is the linear muffin-tin orbitals (LMTOs)\cite{Andersen1975} adopted in early
LDA+DMFT implementations\cite{Anisimov1997,Lichtenstein1998,Savrasov2001a}.  However a more physical choice should be
wannier functions, since the shape of the local orbitals will be
altered in crystals especially when there is strong hybridization
between localized orbital and delocalized $p$-type or $s$-type
orbitals.  The wannier functions are not uniquely determined by the
Bloch wave functions, so different choice can be made, for example,
the Nth-order muffin-tin orbitals\cite{Andersen2000}, the projected wannier functions\cite{Anisimov2005} and
the Maximally localized Wannier functions (MLWFs)\cite{Souza2001,Marzari1997}. Comparisons of
these choices have been made in previous literature. In this paper,
two kinds of local orbitals, atomic like functions and the projected
wannier functions are adopted according to the different systems. 

This implementation of LDA+DMFT method reported in this paper is
developed on an existing DFT package BSTATE (Beijing Simulation Tool
of Atomic TEchnique), which is based on the pseudo-potentials and plane
waves. Unlike the LMTO methods, the local orbitals
do not enter the basis set of plane waves and should be
constructed in a suitable way. Projected wannier functions or MLWFs as local orbitals have been used by previous reported LDA+DMFT implementations based on pseudo-potentials (or projector augmented wave (PAW)) method) and plane wave scheme\cite{Amadon2008b,Trimarchi2008,Korotin2008}. In this report, two kinds of orbitals are used according to the need, one is directly derived from the atomic wave function of an isolated atom and the other is projected wannier functions constructed from these atomic wave functions. In order to be self-contained the construction procedure is presented below in detail.  Since all our calculations are based on plane wave set, the local orbitals will be projected onto these plane waves. 

First we consider the atomic like local orbitals. The localized nature of the correlated bands with $d$ or $f$ characters in crystals ensures that these local orbitals are very similar to the corresponding atomic wave functions of an isolated atom. The atomic wave functions can be picked as the local orbitals directly if the hybridization in crystal could be neglected. All atomic wave functions for the isolated atom could be obtained by the solving the all-electron radical Schr\"{o}dinger equation 
\begin{equation}
  \label{eq:4}
  (T+V(r))\psi_{nl}^{ALL}(r)=E_{nl}\psi_{nl}^{ALL}(r)
\end{equation}
In this manner, all orbitals are well defined by the primer quantum number $n$ and angular quantum number $l$. In realistic systems, the local orbitals usually come from the partially filled $d$ or $f$ shell and could be picked according to its character. Often there is only one type of local orbitals, so the local subspace can be labeled by $l$. Of course, considering the spherical symmetry of the isolated atom,  the spherical harmonics should be multiplied the radical wave functions to form a complete set.  The local orbitals are
\begin{equation}
 \vert\phi_{lm,I}\rangle  =\vert \psi_{l,I}^{ALL}Y_{lm}\rangle,
\end{equation}
and in which $m=1\dots 2l+1$ denote different angular components. 

In methods based on plane waves, it is not desirable to use the
all-electron wave function directly since it requires  a large amount
of plane waves to expand in momentum space. To avoid this problem, in
PAW method all-electron atomic  partial waves  can be used, while in
pseudo-potentials method, pseudo atomic wave functions can be chosen. The latter is used in this paper. During the generation of pseudo-potentials, the Schr\"{o}dinger equation obeyed by the pseudo wave functions is as below
\begin{equation}
  \label{eq:4}
  (T+V^{PS}(r))\psi_{nl}^{PS}(r)=E_{nl}^{PS}\psi_{nl}^{PS}(r)
\end{equation}
 The details on the generation of pseudo-potentials can be referred to previous literature\cite{Vanderbilt1990,Hamann1979}. The spirit of pseudo-potentials is quite straight forward. Beyond a core radius $r_c$, the pseudo-potential $V^{PS}$ play the role of a scattering center just as a real atomic potential $V^{ALL}$ do. Thus, the pseudo eigen energy $E_{nl}^{PS}$ should be the same as the realistic one $E_{nl}^{ALL}$, and both the pseudo wave functions $\vert \psi_{n,I}^{PS}\rangle$  and the pseudo-potential  $V^{PS}$ coincide with the exact wave functions $\vert \psi_{n,I}^{ALL}\rangle$ and the realistic potential $V^{ALL}$ of an isolated atoms beyond this core radius $r_c$, respectively.
\begin{equation}\label{eq:5}
   \eqalign{E_{nl}^{PS}=E_{nl}; \cr
    V^{PS}(r)=V(r), \qquad \psi^{PS}(r)=\psi^{ALL}(r) \qquad r>r_c}
\end{equation}
The quality of pseudo wave functions is the same as the quality of the pseudo-potentials, which could be justified by comparing simple DFT calculations with accepted results. The pseudo wave functions bear the same atomic features as the exact atomic wave functions, and can be picked as the local orbitals as above (angular quantum number $l$ ignored since usually only one kind of local orbital are considered).
\begin{equation}
 \vert\phi_{m,I}\rangle  =\vert \psi_{l,I}^{PS}Y_{lm}\rangle,
\end{equation}
It is convenient to transform the local orbitals from real space to momentum space in crystal calculations
\begin{equation}
  \label{eq:2}
  \vert \phi_{m,{\mathbf k}} \rangle =\sum_{I} e^{i{\mathbf k}\cdot{\mathbf R}_{I}}\vert \phi_{m,I} \rangle .
\end{equation}
Since the atomic wave functions at different site $I$ are not orthogonal, an orthogonalization procedure is needed for the above local orbitals. 

The DFT calculations give a set of Bloch waves $\vert \Psi_{n\mathbf k}\rangle$ spanning the total Hilbert space in which lies the local subspace spanning by the local orbitals. Thus, the physical choice of local orbitals is Wannier functions derived from the Bloch waves. The Wannier functions are localized in real space and could be constructed from the Bloch waves via a unitary transformation. However, this unitary transformation is not unique since the phase factors of Bloch waves are uncertain, which results in that there are many kinds of Wannier functions in use, e.g., the projected Wannier functions, NMTOs and MLWFs. Also it is not necessary to include all the Bloch waves but the relevant bands lying in a certain energy range to construct the Wannier functions. The local orbitals could be picked as a subset of these Wannier functions with specified $d$ or $f$ localized character.

The projecting Wannier functions method is a simple way to construct wannier functions, in which trial wave functions are projecting onto physical relevant Bloch bands. The atomic like local orbitals above are used as trial functions here. The Bloch bands can be selected by band indices ($N_i, ...N_j$)  or by an energy interval ($E_i,E_j$) enclosing them. 
\begin{eqnarray}
  \label{eq:3}
  \vert W_{m,{\mathbf k}}\rangle&=&\sum_{n=N_i}^{N_j} \vert\psi_{n{\mathbf k}}\rangle\langle \psi_{n{\mathbf k}} \vert \phi_{lm,{\mathbf k}}\rangle \nonumber \\
  &=&\sum_{n(E_i<\epsilon_{n{\mathbf k}} <E_j)} \vert\psi_{n{\mathbf k}}\rangle\langle \psi_{n{\mathbf k}} \vert \phi_{m,{\mathbf k}} \rangle .
\end{eqnarray}
These orbitals need normalization and orthogonalization to be true Wannier functions.
\begin{equation}
  \label{eq:6}
    \vert \tilde{W}_{m,{\mathbf k}}\rangle= \sum_{m'} O_{m,m'}({\mathbf k})^{-1/2}\vert {W}_{m,{\mathbf k}}\rangle
\end{equation}
in which $O$ is the overlap matrix between different $  \vert W_{m,{\mathbf k}}\rangle$s
\begin{equation}
  \label{eq:7}
  O_{m,m'}({\mathbf k})=  \langle W_{m,{\mathbf k}}\vert W_{m',{\mathbf k}}\rangle
\end{equation}
Although the construction method can not give the most localized Wannier orbitals, the basic features of the localized bands are captured to a very good extend as indicated by a few reports\cite{Anisimov2005,Amadon2008b,Trimarchi2008,Korotin2008}, and also proved by our calculations in this paper.

\section{LDA+DMFT(Hubbard-I) Formalism}\label{ldadmft}
\begin{figure}
  \begin{center}
  \includegraphics[angle=0,width=0.8\textwidth, height=0.8\textheight, trim=-1.0cm -1.0cm -1.0cm -1.0cm]{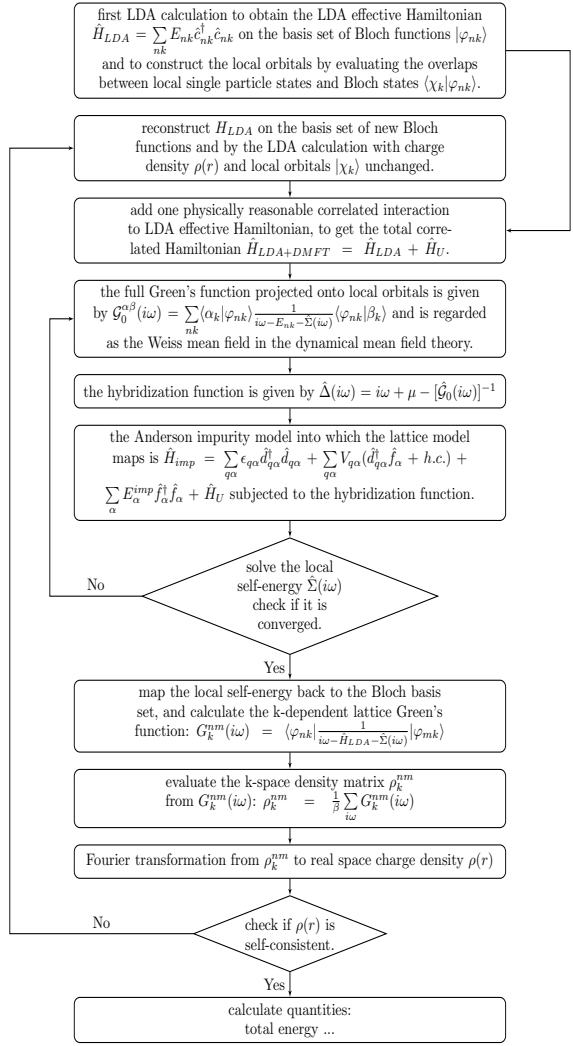}
    \caption{The most general flow chart for LDA+DMFT scheme. The first LDA calculation gives out the band structure on LDA level, and constructs local orbitals, then the full Hamiltonian is established and solved by a DMFT loop, after which the charge density can be recalculated by Fourier transforming the k-space density matrix to real space. The ``non-interacting'' Hamiltonian is thus regenerated based on the new charge density profile and the calculation will be completed when the self consistency has been reached for the full loop.}
    \end{center}
\label{flowchart}
\end{figure}

We will introduce the detail procedge of LDA+DMFT with full loop charge density self consistency in this section. First we plot the general flow
chart in  \fref{flowchart}, which contains a inner DMFT loop and a outer charge density loop. However, if we use Hubbard-I approximation as the impurity solver, the inner DMFT loop is thus neglected.

In the following subsections, we are going to describe the whole process in detail.

\subsection{First LDA Calculation}
The first step of the full loop  LDA+DMFT is an LDA self-consistent calculation, whose main purpose 
is to generate an effective single particle Hamiltonian 
$\hat H_{LDA}$ and construct the correlated orbitals.

Generally the effective LDA Hamiltonian can be expressed as the following:
\begin{eqnarray}\label{H_LDA}
\hat H_{LDA}=\sum_{n\mathbf k}E_{n\mathbf k}\hat c^\dagger_{n\mathbf k}\hat c_{n\mathbf k}
\end{eqnarray}
In the above equation, $n=1\sim N_{band}$ are the joint indices of band and spin;
$E_{n\mathbf k}$ is the eigen energy of the Kohn-Sham equation determined by $\hat{H}_{LDA}|\varphi_{n\mathbf k}\rangle=E_{n\mathbf k}|\varphi_{n\mathbf k}\rangle$,
where $|\varphi_{n\mathbf k}\rangle$ represents a set of orthonormal Bloch functions.

The LDA calculation also complement the construction of local correlated orbitals. 
In the present implementation the atomic orbitals and the Wannier functions are two types of commonly used local basis.
In LDA+DMFT, the local interactions have been considered on the DMFT level only within the local orbitals and in order to set up the DMFT
self consistent equation we need to obtain the overlap matrix between local orbitals and Bloch wave functions, which takes the form of
$S^{\mathbf k}_{\alpha ,n}=\langle\alpha_{\mathbf k}|\varphi_{n\mathbf k}\rangle$, with Greek letter $\alpha$ denote the index of local orbitals.
The completeness of the Bloch basis set gives
\begin{eqnarray}\label{S_norm}
\sum_{n}|S^{\mathbf k}_{\alpha ,n}|^2=1
\end{eqnarray}
for any local orbital $\alpha$ and any k-point.

\subsection{DMFT Loop}
The purpose of the DMFT loop is to calculate the self-energy caused by the local interactions through the DMFT self consistent loop. As we have mentioned before, the DMFT loop is not necessity if we use Hubbard-I approximation as the impurity solver. However, here we first introduce the algorithm with more general impurity solvers.

By adding the local interaction terms to LDA, we get the total Hamiltonian of the system which is to be solved by DMFT,
\begin{eqnarray}\label{H_LDA+DMFT}
\hat H_{LDA+DMFT}=\hat H_{LDA}+\sum_i(\hat H^i_{U}-\hat V^i_{DC})-\mu\hat N
\end{eqnarray}
where $i $ is the site index. For a specified site, we remove the $i$ index and 
\begin{eqnarray}\label{H_U_general}
\hat H_U=\frac{1}{2}\sum_{\alpha\beta\gamma\delta}U_{\alpha\beta\gamma\delta}
\hat f^\dagger_{\alpha}\hat f^\dagger_{\beta}\hat f_{\gamma}\hat f_{\delta}
\end{eqnarray}
is written as a general form of two-body local interactions, in which the creation and annihilation operators of correlated orbitals
$\hat f^\dagger_{\alpha}$ and $\hat f_{\alpha}$ are associated with Bloch operators in 
term of
\begin{eqnarray}\label{f_operation}
\hat f^\dagger_{\alpha}=\sum_{n\mathbf k}\hat c^\dagger_{n\mathbf k}\langle \varphi_{n\mathbf k}|\alpha_{\mathbf k}\rangle\\
\hat f_{\alpha}=\sum_{n\mathbf k}\hat c_{n\mathbf k}\langle\alpha_{\mathbf k}|\varphi_{n\mathbf k}\rangle
\end{eqnarray}

The third term of \eref{H_LDA+DMFT} is the double-counting term correspond to 
the correlated energy that has already been considered in LDA calculation at a Hartree-Fock mean field
level,
\begin{equation}\label{V_DC_LDA}
\hat V_{DC}=\sum_{\alpha}V_{DC}^{\alpha}\hat f^\dagger_{\alpha}\hat f_{\alpha}
\end{equation}   
and this term will be discussed in \sref{section_of_DC}.
The last term of \eref{H_LDA+DMFT}
\begin{equation}\label{N_operator}
\hat N=\sum_{n\mathbf k}\hat c^\dagger_{n\mathbf k}\hat c_{n\mathbf k}
\end{equation}
is the total number of particle operator, while the chemical potential $\mu$ 
controls the occupation number of the unit cell.

\subsubsection{Quantum Impurity Hamiltonian}

In DMFT, the correlation problem on the lattice can be mapped to a quantum impurity model, which contains the same on-site interaction and reads,
\begin{eqnarray}\label{H_imp}
\hat H_{imp}=\sum_{q\alpha}\epsilon_{q\alpha}\hat c^\dagger_{q\alpha}\hat c_{q\alpha}
+\sum_{q\alpha}V_{q\alpha}(\hat c^\dagger_{q\alpha}\hat f_\alpha+h.c.) +\sum_\alpha E^{imp}_\alpha\hat f^\dagger_{\alpha}\hat f_{\alpha}+\hat H_U
\end{eqnarray}
The inference from the rest of the lattice site besides the one considered in the impurity model is simulated by a non-interacting ``heat bath'', which is described by the first term in \eref{H_imp}. The second term describes the coupling between the impurity site and the heat bath and the rest two terms describe the local interactions.

\subsubsection{Hybridization function and Weiss field}
The above quantum impurity model can be solved by the impurity solver, i.e. Hubbard-I, and the self energy $\hat{\Sigma}(i\omega)$ is then 
obtained, with which we can construct the lattice Green's function by applying the same self energy term as,
\begin{equation}\label{dmftloop1}
  [\hat{G}^{\mathbf k}_{lattice}]^{-1}=\rmi \omega-\hat{H}_{\mathbf k}-\hat{\Sigma}_{\mathbf k}(\rmi\omega)+\mu
\end{equation}

The hybridization function, which characterizes the dynamics of the ``heat bath'' is defined as  $\Delta(\rmi\omega)_{\alpha\beta}=\delta_{\alpha\beta}\sum\limits_q\frac{|V_{q\alpha}|^2}{\rmi\omega-\epsilon_{q\alpha}}$ and can be obtained iteratively
by the following DMFT self consistent equation.

\begin{equation}\label{dmftloop2}
  \hat{\Delta}(\rmi\omega)=\rmi\omega-\hat{E}_{imp}-\hat{\Sigma}+\mu-\left[\sum_k G^{\mathbf k}_{lattice}(\rmi\omega)\right]^{-1}
\end{equation}
The above equation is obtained by requiring that the local Green's function on the lattice should equal to
the Green's function of the quantum impurity problem. The equations \eref{dmftloop1} and \eref{dmftloop2} form a closed self consistent loop,
which determines both the self energy and the hybridization function iteratively.

\subsubsection{Double Counting Term}\label{section_of_DC}

In this section we discuss the explicit expression of $E^{imp}_\alpha$ as well as the double-counting term 
in \eref{H_LDA+DMFT}.
 For 3d system under cubic symmetry, the local interaction can be simplified as coulomb interaction $U$ and 
Hund's rule coupling $J$, which can be written as,

\begin{equation}\label{H_U}
\hat{H}_{U}=U\sum_{b}\hat{n}_{b,\uparrow}\hat{n}_{b,\downarrow}
+(U-2J)\sum_{b<b'\atop\sigma\sigma}\hat{n}_{b,\sigma}\hat{n}_{b',\sigma}
-J\sum_{b<b'\atop\sigma}\hat{n}_{b\sigma}\hat{n}_{b'\sigma}
\end{equation}

And the double counting energy in this case has been studied by Held \etal\cite{Held2007}. and  can be approximately chosen as 

\begin{eqnarray}\label{E_DC}
E_{DC}=\frac{1}{2}\bar{U}n_{\mbox{\tiny LDA+DMFT}}(n_{\mbox{\tiny LDA+DMFT}}-1)
\end{eqnarray}
where
\begin{eqnarray}\label{U_eff}
\bar{U}&=&\frac{U+2(M-1)(U-2J)-(M-1)J}{2M-1}\\
M&=&2l+1
\end{eqnarray}
and $n_{\mbox{\tiny LDA+DMFT}}$ is the total number of electrons on correlated orbitals for a specific atom,
hence in \eref{H_LDA} becomes
\begin{eqnarray}\label{V_DC}
\hat V_{DC}&=&\sum_{\alpha}\frac{\partial E_{DC}}{\partial n_\alpha}\hat n_\alpha\nonumber\\
&=&\sum_{\alpha}\bar{U}(n_{\mbox{\tiny LDA+DMFT}}-\frac{1}{2})\hat n_\alpha
\end{eqnarray}
The final expression of $E^{imp}$ is
\begin{eqnarray}\label{E^imp}
E^{imp}_\alpha&=&\sum_k \langle\alpha_k|\hat H_{LDA}|\alpha_k\rangle-V_{DC}^{\alpha}
\end{eqnarray}
where
\begin{eqnarray}\label{V_DC_component}
V_{DC}^{\alpha}=\bar{U}(n_{\mbox{\tiny LDA+DMFT}}-\frac{1}{2})
\end{eqnarray}

The way to remove the form of double-counting energy is not unique, 
and in fact this process needs intuition of physics. The double-counting
discussed in earlier sections is a very usual way which is also used in LDA+$U$
method. Besides, the following two ways are also commonly used
\begin{itemize}
\item to remove self-energy at zero-frequency $\hat\Sigma(\rmi\omega)\rightarrow\hat\Sigma(\rmi\omega)-\hat\Sigma(0)$
\item to remove self-energy at infinity-frequency $\hat\Sigma(\rmi\omega)\rightarrow\hat\Sigma(\rmi\omega)-\hat\Sigma(\infty)$
\end{itemize}
Furthermore, the double-counting energy is also regarded as ``impurity solver dependent''. For example,
it is reasonable to use an integer to replace $n_{\mbox{\tiny LDA+DMFT}}$ in \eref{E_DC} for the Hubbard-I solver\cite{Pourovskii2007}, and in this paper we also follow their suggestion.

\subsubsection{Hubbard-I solver}

The essential point of Hubbard-I solver is to neglect the effect of the heat bath and take the atomic self energy
as the zeroth order approximation for the quantum impurity problem, which can be written as
\begin{eqnarray}\label{Dyson}
\hat\Sigma^{atom}=[\rmi\omega_n+\mu_{atom}-\hat E^{imp}]^{-1}-[\hat G^{atom}]^{-1}
\end{eqnarray}
In the above equation $\hat G^{atom}$ is the Green's function for a single atom, which can be expressed as
\begin{eqnarray}\label{G_imp_atom}
G^{atom}_{\alpha\alpha'}(\rmi\omega)=\sum_{\Gamma\Gamma'}
\frac{(F^\alpha)_{\Gamma\Gamma'}(F^{\alpha'\dagger})_{\Gamma'\Gamma}(X_{\Gamma}+X_{\Gamma'})}
{\rmi\omega-E_{\Gamma'}+E_{\Gamma}}
\end{eqnarray}
where $|\Gamma\rangle$ and $|\Gamma'\rangle$ are the atomic eigenstates obtained by exact diagonalization of a single atom problem,
$X_{\Gamma'}={e^{-\beta E_{\Gamma'}}}/({\sum_\Gamma e^{-\beta E_\Gamma}})$
represents the occupation probability of the local configuration $|\Gamma\rangle$, 
and $F^{\alpha}_{\Gamma\Gamma'}=\langle\Gamma|f_{\alpha}|\Gamma'\rangle$.

Hubbard-I approximation is 
\begin{eqnarray}\label{Hub1_approx}
\hat\Sigma\approx\hat\Sigma^{atom}
\end{eqnarray}

In reference \cite{Savrasov2006a} , Savrasov \etal propose that the above self energy can be written by the summation of a set of
poles, which greatly simplifies the DMFT calculation, because it is not necessary to handle the full frequency dependence
of the Green's function. In Hubbard-I approximation, it can be  proved that both the atomic Green's function and
self energy are diagonal and can be written in the form of pole expansion if the following two necessary conditions
are satisfied. 1) The single particle basis used here should be the one which diagonalize the local density matrix;
2) The atomic Hamiltonian only contains the two-body interaction terms. Obviously by chosen the proper single 
particle basis the atomic Hamiltonian \eref{H_imp} in general will satisfy the above two condition and thus can be always
written in terms of pole expansion. Unlike the reference\cite{Savrasov2006a}, where they only use a few poles to capture the main features of the self energy, in the present implementation, we keep all the poles in the self energy, which 
makes it more accurate. From \eref{G_imp_atom}, it is very clear that the atomic green's function has already 
been expressed in terms of poles and the pole expansion of the corresponding self energy can be obtained using 
\eref{Dyson}, which is introduced in details in the Appendix. Therefore in general the above atomic self energy can
be written as
\begin{equation}\label{se_pole}
  \hat{\Sigma}(\rmi\omega)=\hat{\Sigma}({\infty}) + \sum^{np}_{i=1} \hat{V}^\dag_i (\rmi\omega-p_i)^{-1} \hat{V}_i
\end{equation}
where $i$ labels the number of poles, $np$ is the total number of poles in the self energy 
and $\hat{V}_i$ is a vector defined in the orbital space describing the
distribution of the $ith$ ``pole states'' among the local orbitals.

\subsection{Correction of the Density Matrices and Pole Expansion of the Self-energy}
Once we obtain the converged local self-energy, we are at the point to correct the k-dependent lattice Green's functions as well as the density matrices. In general we need to do the summation over the Matsubara frequencies, which is time consuming for realistic materials contains lots of 
bands. But it can be greatly simplified when the self energy can be expressed as the summation of poles, which can be written as

\begin{eqnarray}\label{rho_LDA+DMFT}
&&\langle\varphi_{n_1\mathbf k}|\hat\rho_{LDA+DMFT}|\varphi_{n_2\mathbf k}\rangle \nonumber \\
&=&\frac{1}{\beta}\sum_{i\omega_n}
\langle\varphi_{n_1\mathbf k}|\frac{1}{\rmi\omega_n+\mu-\hat H_{LDA}+\hat V_{DC}-\hat\Sigma(i\omega_n)}|\varphi_{n_2\mathbf k}\rangle
\end{eqnarray}

As pointed out firstly in reference \cite{Savrasov2006a}, when the self energy can be written in terms of  poles, the full green's
function can be expressed as the ``physical'' part of an enlarged ``pseudo Hamiltonian'', defined as 

\begin{eqnarray}\label{pseudo_H_k}
\hat H_{ps}(\mathbf k)=
\left(
\begin{array}{cc}
 \hat H_{LDA}(\mathbf k)-\hat V_{DC}+\hat\Sigma(\infty)& \qquad-\hat V^\dagger  \\
 -\hat V & \qquad\hat P
\end{array}
\right)
\end{eqnarray}
where $\hat H_{LDA}(\mathbf k)=E_{n\mathbf k}\hat c^\dagger_{n\mathbf k}\hat c_{n\mathbf k}$ is the physical Hilbert space, $\hat {V}_i$ is defined in 
\eref{se_pole} and $\hat P=Diag(p_1,p_2,....,p_{np})$. Therefore the physical green's function can be expressed
in terms of eigenstate of  the pseudo Hamiltonian simply as
\begin{eqnarray}
  & &\langle\varphi_{n_1\mathbf k}\bigg|\frac{1}{\rmi\omega_n-\hat H_{LDA}(\mathbf k)+\hat V_{DC}-\hat\Sigma(i\omega_n)}\bigg|\varphi_{n_2\mathbf k}\rangle\nonumber \\
  &= & \langle\varphi_{n_1\mathbf k} | \psi^{ps}_{l\mathbf k}\rangle \frac{1}{\rmi\omega_n-E^{ps}_l(\mathbf k)}\langle \psi^{ps}_{l\mathbf k}|\varphi_{n_2\mathbf k}\rangle 
\end{eqnarray}

where $|\psi^{ps}_{l\mathbf k}\rangle$ and $E^{ps}_{l}(\mathbf k)$ are the eigenstate and eigenvalue of pseudo Hamiltonian respectively.

Then the sum of frequencies in \eref{rho_LDA+DMFT} can be performed directly
\begin{eqnarray}\label{rho_LDA+DMFT_componet_final}
&&\langle\varphi_{n_1\mathbf k}|\hat\rho_{LDA+DMFT}|\varphi_{n_2\mathbf k}\rangle\nonumber \\
&=&\frac{1}{\beta}\sum_{i\omega_n}\sum_{l\mathbf k}
\langle\varphi_{n_1\mathbf k}|\psi^{ps}_{l\mathbf k}\rangle\frac{1}{\rmi\omega_n-E_{l}^{ps}(\mathbf k)}\langle\psi^{ps}_{l\mathbf k}|\varphi_{n_2\mathbf k}\rangle\nonumber \\
&=&\sum_{l\mathbf k}\langle\varphi_{n_1\mathbf k}|\psi^{ps}_{l\mathbf k}\rangle\langle\psi^{ps}_{l\mathbf k}|\varphi_{n_2\mathbf k}\rangle
n_F[E_{l}^{ps}(\mathbf k)-\mu]
\end{eqnarray}
in which $\mu$ is exactly the chemical potential in \eref{H_LDA+DMFT}, determined by the occupation number
of electrons $N_{tot}$ in the unit cell
\begin{eqnarray}\label{fix_mu}
N_{tot}&=&\sum_{n\mathbf k}\langle\varphi_{n\mathbf k}|\hat\rho|\varphi_{n\mathbf k}\rangle\nonumber \\
&=&\sum_{ln\mathbf k}|\langle\varphi_{n\mathbf k}|\psi^{ps}_{l\mathbf k}\rangle|^2
n_F[E_{l}^{ps}(\mathbf k)-\mu]
\end{eqnarray}

\subsection{Evaluation of Real Space Charge Density and Complete of the Full Loop}
After DMFT obtains the corrected density matrix $\hat\rho$, the real space charge density will be generated based on it
by Fourier transformation, and then, the
new LDA Hamiltonian as well as the new overlap matrices between Bloch states and local orbitals will be recalculated, which
closes the full iteration loop for the charge density self consistency. 
The occupation number of the Bloch basis can also be obtained by, 
\begin{eqnarray}\label{nf_LDA+DMFT}
n_{\mbox{\tiny LDA+DMFT}}=\sum_{\mathbf k,n_1n_2,\alpha}
\langle\alpha_{\mathbf k}|\varphi_{n_1\mathbf k}\rangle
\langle\varphi_{n_1\mathbf k}|\hat\rho|\varphi_{n_2\mathbf k}\rangle
\langle\varphi_{n_2\mathbf k}|\alpha_{\mathbf k}\rangle
\end{eqnarray}

\subsection{Calculation of Physical Quantities}
\subsubsection{Energy Functional}
Most of the  physical quantities we are interested in can be calculated once we have reached the charge density self consistency.
First we present the  
energy functional of the full loop LDA+DMFT, which can be used to calculate many quantities, i.e. the total energy, force and so on.
It can be written as,
\cite{Pourovskii2007}
\begin{eqnarray}\label{E_total}
E=E_{LDA}[\rho]-\langle H_{KS}\rangle_{LDA}+\langle H_{KS}\rangle
+\langle H_U\rangle-E_{DC}
\end{eqnarray}
in which $E_{LDA}[\rho]$ is the expression of the energy within density-functional theory;
$\langle H_{KS}\rangle_{LDA}=\tr[\hat H_{LDA}\hat G_{LDA}]$ is the non-interacting energy 
at LDA level; $\langle H_{KS}\rangle=\tr[\hat H_{LDA}\hat G]$ is the non-interacting energy at DMFT level; 
$\langle H_U\rangle=\frac{1}{2}\tr[\hat\Sigma\hat G]$ 
is the interaction energy caused by local correlation interactions; the 
last term $E_{DC}$ is the double counting energy given before in \eref{E_DC}.
In above the meaning for ``trace'' is defined as
\begin{eqnarray}\label{trace}
\tr[\mathcal{A}]=\frac{1}{\beta}\sum_{n\mathbf k}\sum_{\rmi\omega_n}\langle\varphi_{n\mathbf k}|\mathcal{A}(\rmi\omega_n)|\varphi_{n\mathbf k}\rangle
\end{eqnarray}
and the integral path surrounds all the energies of occupation states. 

The first and the second term is easily calculated in the LDA framework, and the three
terms remaining must be evaluated in the DMFT process, explicitly
\begin{eqnarray}\label{H_KS}
\langle H_{KS}\rangle=\sum_{n\mathbf k}E_{n\mathbf k}\langle\varphi_{n\mathbf k}|\hat\rho|\varphi_{n\mathbf k}\rangle
\end{eqnarray}
and
\begin{eqnarray}\label{H_U_energy}
\langle H_U\rangle&=&\frac{1}{2}\tr[\hat\Sigma\hat G]\nonumber \\
&=&\frac{1}{2}\tr[(\hat G^{-1}_0-\hat G^{-1})\hat G]\nonumber \\
&=& \frac{1}{2} \left [\sum_{m,n\mathbf k}(E^{ps}_{m\mathbf k}-\mu)|\langle \varphi_{n\mathbf k}|\psi_{m\mathbf k}\rangle|^2n_F(E^{ps}_{m\mathbf k}-\mu) \right. \nonumber \\
& \qquad &\left. -\langle H_{KS}\rangle+\langle\hat{V}_{DC}\rangle\right]
\end{eqnarray}
while $\langle H_{KS}\rangle$ is expressed in \eref{H_KS}, and
$\langle\hat V_{DC}\rangle$ is the expected value of $\hat V_{DC}$ at DMFT level
\begin{eqnarray}\label{<V_DC>}
\langle\hat V_{DC}\rangle&=&\sum_i
\bar{U}_i(n^i_{\mbox{\tiny LDA+DMFT}}-\frac{1}{2})\sum_{\alpha}\langle\hat n^i_\alpha\rangle\\
&=&\bar{U}_i(n^i_{\mbox{\tiny LDA+DMFT}}-\frac{1}{2})n^i_{\mbox{\tiny LDA+DMFT}}
\end{eqnarray}
Notice that here emerges the site index $i$. We should be very careful in the case that 
the number of correlated atoms is larger than one. The final expression of the total energy 
is
\begin{eqnarray}\label{E_total_final_expression}
\fl E=E_{LDA}[\rho]-\langle H_{KS}\rangle_{LDA}+\langle H_{KS}\rangle \nonumber \\
+\frac{1}{2}\left[\sum_{m,nk}(E^{ps}_{mk}-\mu)|\langle \varphi_{nk}|\psi_{mk}\rangle|^2n_F(E^{ps}_{mk}-\mu)-\langle H_{KS}\rangle\right] \nonumber \\
+\frac{1}{4}\sum_i\bar U_i
n^i_{\mbox{\tiny LDA+DMFT}}
\end{eqnarray}

However, if we use Hubbard-I as the impurity solver, and remove double-counting term by using an integer, the energy functional is given by Haule\cite{Haule2010}
\begin{eqnarray}\label{E_total_final_expression}
\fl E=E_{LDA}[\rho]-\langle H_{KS}\rangle_{LDA}+\langle H_{KS}\rangle \nonumber \\
+\frac{1}{2}\left[\sum_{m,n\mathbf k}(E^{ps}_{m\mathbf k}-\mu)|\langle \varphi_{n\mathbf k}|\psi_{m\mathbf k}\rangle|^2n_F(E^{ps}_{m\mathbf k}-\mu)-\langle H_{KS}\rangle\right] \nonumber \\
+\frac{1}{4}\sum_i\bar U_i
n^i_0
\end{eqnarray}
where $n^i_0$ refers to the integer used to remove double counting for the $i$th correlated atom.

\subsubsection{Expressions of DOS and PDOS}
In order to calculate the density of state (DOS) or the partial density of state (PDOS)
on correlated orbitals, it is necessary to calculate the retarded form of the Green's function.
By using the virtue of pole expansion method, we write directly
\begin{eqnarray}\label{DOS}
D(\epsilon)=(-\frac{1}{\pi})\mathrm{Im}\bigg[\sum_{\mathbf k,m,n}
\frac{\langle\varphi_{n\mathbf k}|\psi_{m\mathbf k}\rangle\langle\psi_{m\mathbf k}|\varphi_{n\mathbf k}\rangle}
{\epsilon+\mu-E^{ps}_{m\mathbf k}+\rmi\eta}\bigg]\ (\eta \rightarrow 0^+)
\end{eqnarray}
and the PDOS of orbital $\alpha$
\begin{eqnarray}\label{PDOS}
D_\alpha(\epsilon)&=&(-\frac{1}{\pi})\mathrm{Im}\left[\sum_{\mathbf k,m}
\frac{\langle\alpha_{\mathbf k}|\psi_{m\mathbf k}\rangle\langle\psi_{m\mathbf k}|\alpha_{\mathbf k}\rangle}
{\epsilon+\mu-E^{ps}_{m\mathbf k}+\rmi\eta}\right]\ (\eta \rightarrow 0^+)
\end{eqnarray}
where
\begin{eqnarray}
\langle\alpha_{\mathbf k}|\psi_{m\mathbf k}\rangle=\sum_{n}\langle\alpha_{\mathbf k}|\varphi_{n\mathbf k}\rangle
\langle\varphi_{n\mathbf k}|\psi_{m\mathbf k}\rangle
\end{eqnarray}


\section{Benchmark}\label{benchmark}
To validate the LDA+DMFT framework based on pseudo-potentials-planewave package and pole expansion of self energy, we benchmark our implementation by applying to $\gamma$-cerium, americium and paramagnetic NiO. These three are typical strongly correlated systems in which the valence electrons are believed to be on the localized side as reported in previous literatures. We  use the Hubbard-I method as impurity solver, which is good enough to capture the atomic-like features in the Mott insulators, so we expect that these systems could be well described by our method. 
We would emphasize here that in our implementation we can replace the Hubbard-I solver by any of the solvers as long as the self energy can be written in pole expansion form.

\subsection{Cerium}
The cerium metal attracts lots of research interests for its isostructural volume-collapse transition from $\gamma$ phase to $\alpha$ phase.  The volume change is about $15\%$ during the transition, which is possibly driven by the entropy change{\cite{Amadon2006}}.  Normal LDA calculations could not give a correct description of cerium, as show in \tref{ce_lattice}, the equillibrium volume of cerium given by LDA calculations is smaller than the experimental volume of $\alpha$ phase. This is due to the fact that in LDA  the $f$-electrons are treated as itinerant and the strong 
correlation effects among them  can not be well captured.
When $f$-electron is treated as core electrons, we see that the equilibrium volume is larger but close to $\gamma$ phase. These simple LDA results 
implies that in $\gamma$ phase, the $f$-electrons are more localized than in $\alpha$-phase. Therefore $f$-electrons in $\gamma$ phase is quite
close to the localized picture like the situation in Mott insulators, which makes it suitable to applying the Hubbard-I solver.
The $\alpha$-Ce which is stable in low temperature is a correlated metal, which is confirmed by many other LDA+DMFT calculations\cite{Amadon2006}, and also by our newly development LDA+Gutzwiller method\cite{Tian2011}.

In our calculation for $\gamma$-Ce, we used ultrasoft pseudo-potentional in which 4\emph{f}, 5\emph{s}, 5\emph{d}, 6\emph{s} orbitals are taken as valence orbitals. The energy cutoff for plane-wave expansion is 12.5 Ha for convergence. Calculations are performed with $10\times 10\times 10$ k-points. For DMFT calculation, the on-site Coulomb interaction $U$ is chosen to be 6.0 eV as suggested in previous reports.\cite{Haule2005,Amadon2006,Amadon2011}

\begin{figure}[hbp]
  \begin{center}
  \includegraphics[angle=270,width=0.8\textwidth]{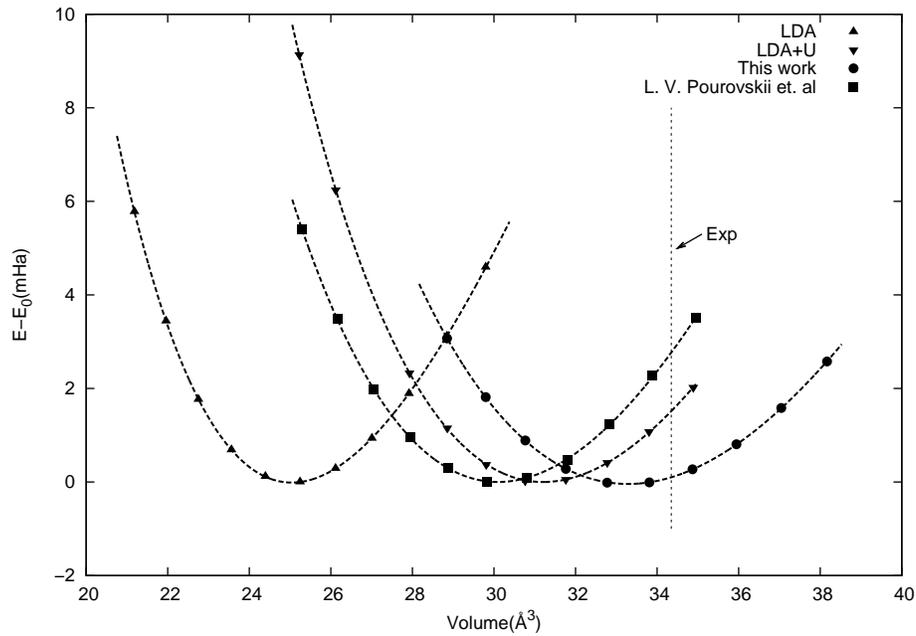}
  \end{center}
  \caption{Total energy versus volume curves obtained by LDA, LDA+$U$\cite{Pourovskii2007}, and LDA+DMFT. The dashed lines are fitted by Birch EOS. The energy were shifted for different curves. The experimental equilibrium volume is from reference \cite{Jeong2004}}
\label{ce_ev}
\end{figure}

In \fref{ce_ev} the total energy versus volume for $\gamma$-cerium obtained by LDA, LDA+$U$ and LDA+DMFT are plotted. The
equilibrium volume and bulk modulus are also shown in \tref{ce_lattice}.  We can find in \fref{ce_ev} that the equilibrium volume
obtained by our LDA+DMFT calculation is
very close to the experimental data, which shows great improvements over LDA . 
The bulk modulus is also improved  but still larger than the experimental data, which may imply the contribution for lattice vibration
is inneglectable  as discussed in reference \cite{Pourovskii2007}.
\begin{table}[hbp]
  \caption{\label{ce_lattice} Lattice parameter and bulk modulus of $\gamma$-cerium according to both experiment and calculations}
  \begin{indented}
    \lineup
    \item[]\begin{tabular}{@{}lll}
      \br
      & Volume($\mathrm{\AA}^3$)  & Bulk modulus(GPa) \\
      \mr
      Experiments & 34.35\cite{Jeong2004}   & 21\cite{Decremps2009}  \\
      LDA      & 22.36  & 62.09  \\
      LDA(fcore)\cite{Huang2007} & 36.39 & 30.2 \\
      LDA+$U$/PAW\cite{Amadon2008a}   & 32.0  & 34  \\
      LDA+DMFT\cite{Pourovskii2007}& 30.10 & 48.49  \\
      LDA+DMFT & 33.29   & 38.27  \\
      \br
    \end{tabular}
    \end{indented}
\end{table}

\begin{figure}
  \begin{center}
  \includegraphics[angle=270,width=0.8\textwidth]{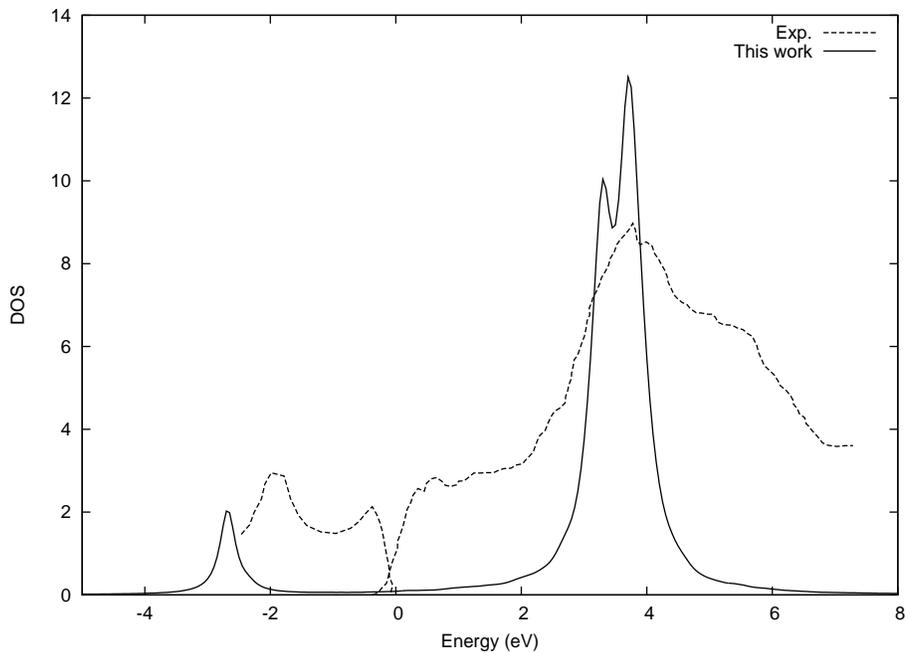}
  \end{center}
  \caption{4\emph{f} pdos for $\gamma$ cerium in LDA+DMFT (Hubbard I), XPS and BIS experimental data are from reference \cite{Wuilloud:1983p2403} and \cite{Wieliczka1982}}
  \label{ce_pdos}
\end{figure}

\Fref{ce_pdos} shows the 4\emph{f} partial DOS for $\gamma$ cerium at the equilibrium volume, the position of the two Hubbard bands agree well with the experimental result.

\subsection{Americium}
Americium, which is widely used in smoke detectors, can be viewed as the "changing point" of the actinide series, where the 5\emph{f} electrons 
changes from delocalized to localized states\cite{Moore2009}. Valence-band ultraviolet photoemission experiment by Naegele \etal\cite{Naegele1984} shows that the 5\emph{f} states are strongly localized. Am undergoes a series of structure phase transitions with the increment of 
pressure, from double hexagonal close packed(dhcp, P6$_3$/mmc), to face centered cubic(fcc, Fm3m, 6.1 GPa), face centered orthorhombic(Fddd, 10.0 GPa), and to primitive orthorhombic structure(Pnma, 16 PGa)\cite{Heathman2000}. 
Therefore Am provides an interesting playfield to investigate the relation between $f$-electron localization and structure transition. Here we focus on the fcc phase and dhcp phase.

In the calculation, we used ultrasoft pseudo-potentional contains 5\emph{f}, 6\emph{p}, 6\emph{d}, 7\emph{s} orbitals as valence orbitals, the plane-wave expansion kinetic energy cutoff of 12.5 Ha. The LDA self-consistent calculations were performed with $10\times 10\times 10$ k-points grid for fcc phase, $9\times 9\times 3$ k-points grid for dhcp phase. For the volume calculation, we only considered density-density interaction, and on-site Coulomb interaction $F^{(0)}=4.5$ eV, for the partial DOS calculation, we considered general interaction, and take the atomic values $F^{(2)}=7.2$ eV, $F^{(4)}=4.8$ eV, $F^{(6)}=3.6$ eV\cite{Carnall1964}.

\begin{figure}[hbp]
  \begin{center}
  \includegraphics[angle=270,width=0.8\textwidth]{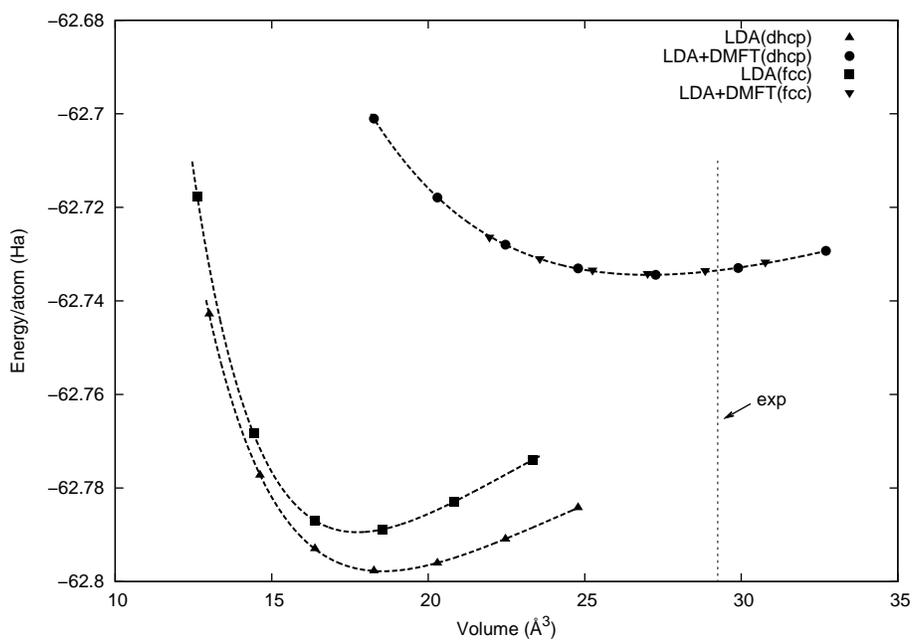}
  \end{center}
  \caption{Total energy per atom of Am obtained by LDA and LDA+DMFT. The LDA+DMFT results of Am for both phase are very close, and the experimental equilibrium volume is denoted by dot line.}
  \label{am_ev}
\end{figure}

\begin{figure}[hbp]
  \begin{center}
  \includegraphics[angle=270,width=0.8\textwidth]{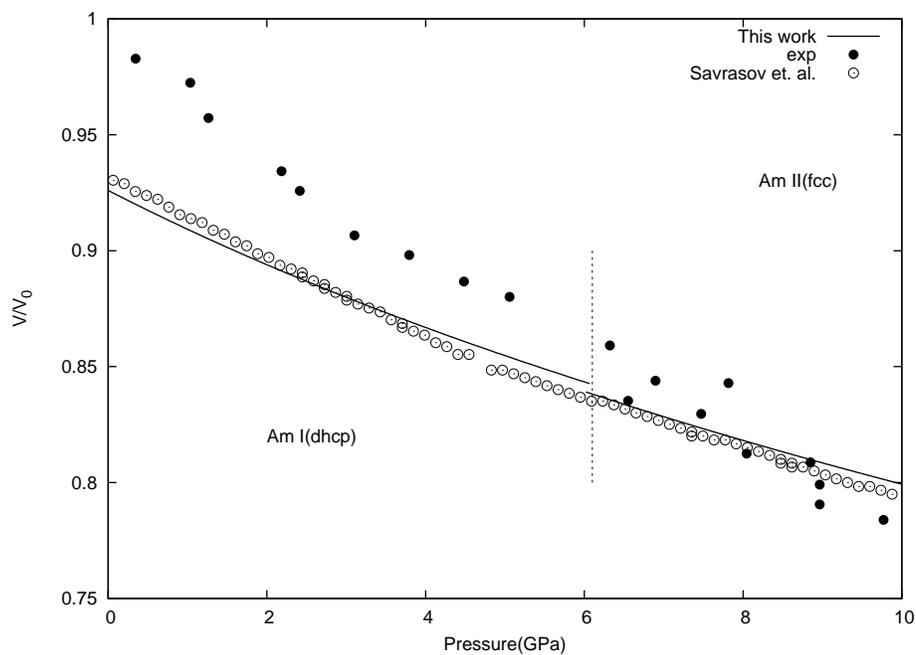}
  \end{center}
  \caption{Calculated P-V relation of dhcp and fcc Am. Here, $V_0$ is experimental equilibrium volume, The experimental P-V results\cite{Lindbaum2001} are shown by black dots.}
  \label{am_pv}
\end{figure}

We present the optimized volume and bulk modulus for both dhcp and fcc phase in \tref{am_lattice} and \fref{am_ev}. 
Our LDA+DMFT results are quite close to the experimental data and show great improvement over LDA.
Also our results are quite consistent with the reference \cite{Savrasov2006a}, in which the full potential LMTO methods
are used in the LDA part.  The equation of state of Am obtained by our LDA+DMFT calculation is plotted in \fref{am_pv},
which again shows very good agreement with both the experimental data and previous LDA+DMFT calculation based
on LMTO method\cite{Savrasov2006a}. 

\begin{table}[hbp]
    \caption{\label{am_lattice}Am equilibrium volume and bulk modulus for dhcp and fcc phase by different methods, together with the experimental results\cite{Heathman2000,Lindbaum2001}.}
    \begin{indented}
      \lineup
      \item[]\begin{tabular}{@{}lll}
        \br
        & Volume($\mathrm{\AA}^3$)  & Bulk modulus(GPa)  \\
        \mr
        experiment    & 29.250 & 29.9  \\
        GGA(dhcp)\cite{Kutepov2004}      & 19.916  &  70.0   \\
        LDA(dhcp) & 18.55 &  118.99  \\
        LDA(fcc) & 17.73 &  169.40 \\
        LDA+DMFT(dhcp) & 27.04  &  52.70  \\
        LDA+DMFT(fcc) & 26.99  & 50.75 \\
        LDA+DMFT\cite{Savrasov2006a}& 27.4  & 45  \\
        \br
      \end{tabular}
    \end{indented}
\end{table}

\begin{figure}[hbp]
  \begin{center}
  \includegraphics[angle=270,width=0.8\textwidth]{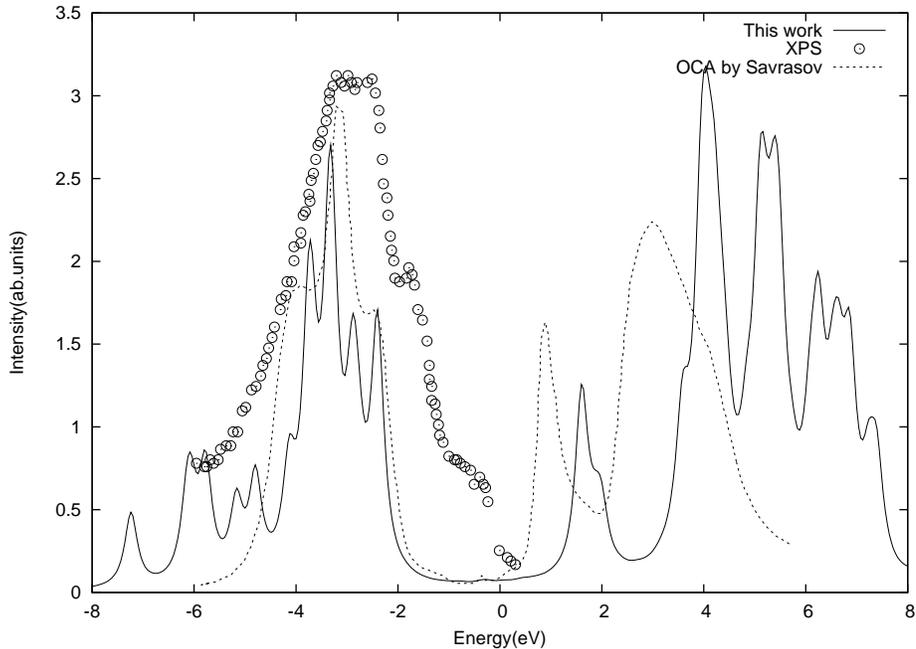}
  \end{center}
  \caption{The spectral function of Am evaluated by LDA+DMFT. The experimental photoemission results is from reference \cite{Naegele1984} are denoted by solid dots.}
  \label{am_dos} 
\end{figure}

\Fref{am_dos} shows the 5\emph{f} partial DOS of fcc phase Am obtained both by our LDA+DMFT calculation and LDA+DMFT(OCA) by Savrasov\cite{Savrasov2006a},together with the photoemission data from ref.\cite{Naegele1984}. Our results is quite consistent with the
 previous calculations and the experimental results.


\subsection{Paramagnetic phase of NiO}
\begin{figure}
  \begin{center}
\includegraphics[width=0.8\textwidth]{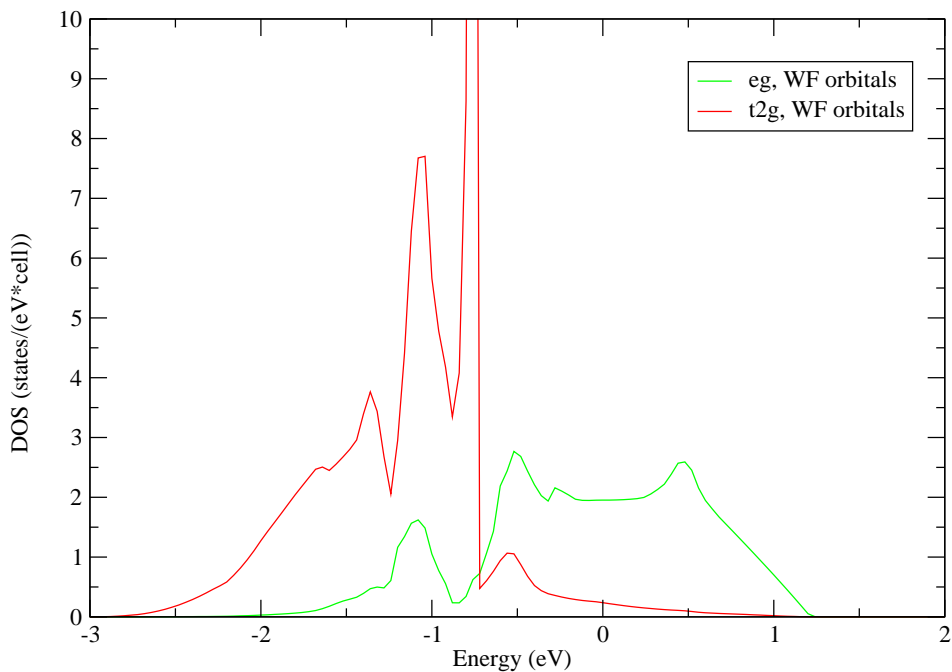}
\end{center}
\caption{The partial DOS of NiO obtained by LDA. The main contribution to the Fermi surface is attributed to eg-like Wannier orbitals, and the t2g-like orbitals are almost fully occupied.} 
\label{NiO_LDA_WF_PDOS}
\end{figure}

\begin{figure}
  \begin{center}
\includegraphics[width=0.8\textwidth,trim=-1.5cm -1.5cm -1.5cm -1.5cm]{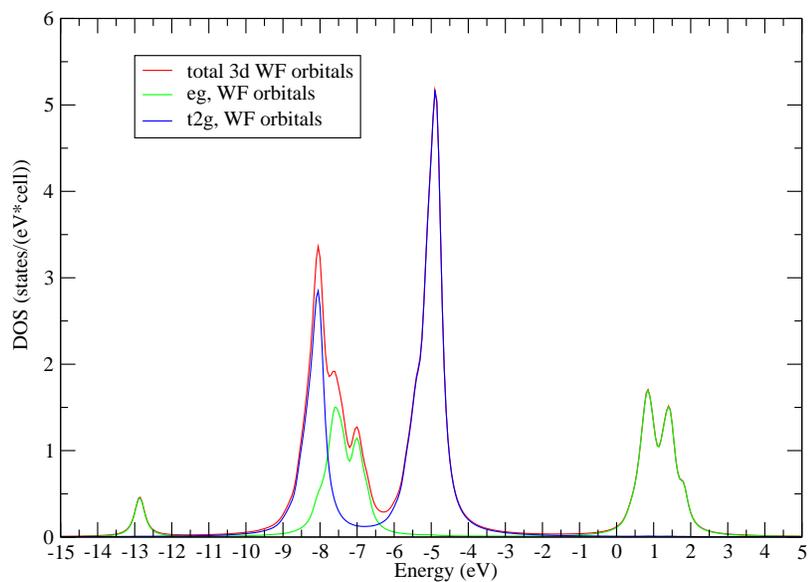}
\end{center}
\caption{The spectral function of NiO obtained by LDA+DMFT calculation.  As shown in the figure, compared with LDA results, the t2g-like orbitals are still fully occupied while the eg-like orbitals split into upper and lower Hubbard bands.} 
\label{NiO_LDA+DMFT_WF_PDOS}
\end{figure}

NiO has been heavily studied as a prototype of Mott insulators. It has been studied within the frame of LDA+DMFT by many groups\cite{Ren2006,Kunes2007}. 
Therefore NiO can  be used as a good benchmark material for the implementation of LDA+DMFT.
In this paper, we apply our code to study the paramagnetic phase of NiO. First we plot the partial density of states (PDOS) obtained
by LDA in \fref{NiO_LDA_WF_PDOS}, which clearly shows metallic behavior and is not consistent with the experiments. In our LDA+DMFT(Hub1)
calculation, we use the wannier functions as the local basis and choose on-site interaction $U=8.0eV$, Hund's coupling $J=1.0eV$.
The Mott insulating features of NiO then can be well captured by our LDA+DMFT method, as shown by the PDOS
in \fref{NiO_LDA+DMFT_WF_PDOS}). 

\begin{figure}
  \begin{center}
\includegraphics[width=0.8\textwidth]{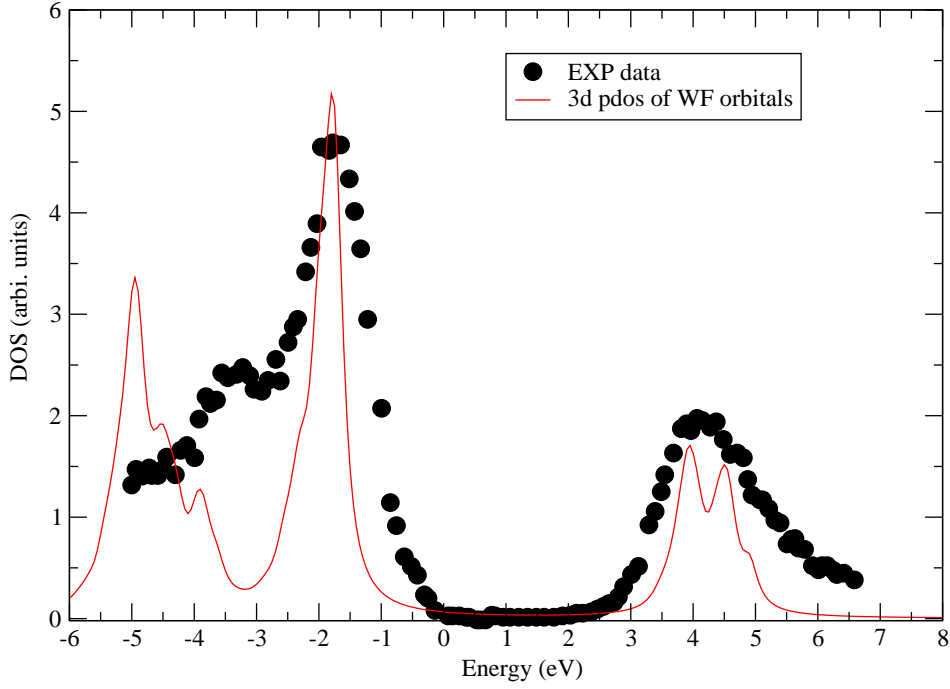}
\end{center}
\caption{The spectral function of NiO evaluated by LDA+DMFT, with the comparison of experimental data\cite{Sawatzky1984}. The behavior of the density of state near the fermi surface fits well with the experimental data.} 
\label{NiO_exp_VS_pdos}
\end{figure}

The energy gap obtained by our calculation is around 4.0eV, which is also quite consistent with the experiments. The photo emission and BIS data obtained by Sawatzky and Allen\cite{Sawatzky1984} are also plotted in \fref{NiO_exp_VS_pdos}. We can find that our results fit the photo emission and BIS data very well.

\section{Conclusions}\label{conclusions}
The new implementation of LDA+DMFT based on the pseudo-potential plane-wave method is introduced in detail
in this paper. We choose the Hubbard-I method as the impurity solver to solve the quantum impurity problem generated by DMFT, which is quite suitable for the Mott insulator materials. The most important advantage of Hubbard-I method is that it is simple enough, which makes the full loop charge self consistent calculation accessible. We also point out in the paper that the difficulty of handling frequency dependent Green's function can be completely removed by expressing the self energy in terms of pole expansion, which greatly raise the efficiency of the method. Finally we benchmark our implementation of LDA+DMFT on several important correlation materials including $\gamma$-Ce,Am and NiO. Our results for all these materials fit very well with both the experimental data and the previous LDA+DMFT results.

\appendix
\section{Self-energy in pole expansion form}\label{appendix}
In this appendix, we will introduce how to evaluate self-energy from Green's function in the pole expansion form.

Green's function can be expressed as
\begin{equation}
  G(i\omega)=\sum^{N_G}_{i=1}\frac{V_i}{i\omega-P_i}
\end{equation}
where $P_i$ is the $i$th pole of the Green's function, $V_i$ is the weight of the pole, and $N_G$ is the number of Green's function poles. Besides the self-energy can be expressed as
\begin{equation}
  \Sigma(i\omega)=\sum^{N_S}_{i=1}\frac{W_i}{i\omega-Q_i}+\Sigma(\infty)
\end{equation}
where $Q_i$ is the $i$th pole of the self-energy, $W_i$ is the weight of the pole, and $N_S$ is the number of self-energy poles.

Then the Dyson's equation \eref{Dyson} becomes
\begin{equation}\label{dyson_pole}
  \sum^{N_S}_{i=1}\frac{W_i}{i\omega-Q_i}+\Sigma(\infty)=i\omega+\mu-\epsilon_{imp}- \left[\sum^{N_G}_{i=1}\frac{V_i}{i\omega-P_i} \right]^{-1}
\end{equation}

First consider the $\Sigma(\infty)$, when $\omega \rightarrow \infty$
\begin{eqnarray}
  \Sigma(\omega\rightarrow\infty) &= i\omega + \mu -\epsilon_{imp}-\left[ \sum^{N_G}_{i=1}\frac{V_i}{i\omega-P_i}\right]^{-1} \nonumber \\
  &= i\omega +\mu-\epsilon_{imp}-\left[\frac{1}{i\omega}\sum^{N_G}_{i=1}\frac{V_i}{1-\frac{P_i}{i\omega}} \right]^{-1} \nonumber \\
  &=i\omega+\mu-\epsilon_{imp}-i\omega \left[ \sum^{N_G}_{i=1}V_i -\frac{1}{i\omega}\sum^{N_G}_{i=1} V_i P_i \right]^{-1} \nonumber \\
  &=i\omega + \mu-\epsilon_{imp}-i\omega\left( 1-\frac{\sum^{N_G}_{i=1}V_i P_i}{i\omega}\right) \nonumber \\
  &=\mu -\epsilon_{imp}+\sum^{N_G}_{i=1} V_iP_i
\end{eqnarray}

For the the self energy poles in \eref{dyson_pole}, it corresponding to the zero in the square bracket of the right side of \eref{dyson_pole}, that is for $ Q \in \{ Q_i\}$, we have
\begin{equation}
  \sum^{N_G}_{i=1}\frac{V_i}{Q-P_i}=0
\end{equation}
The left hand side is monotonically decreasing between two adjacent Green's function poles, so we could use bisection method to find the self energy poles.

We rewrite \eref{dyson_pole} as
\begin{equation}
  \sum^{N_s}_{i=1}\frac{W_i}{i\omega-Q_i}=i\omega+\mu-\epsilon-\Sigma(\infty)-\left[\sum^{N_g}_{i=1}\frac{V_i}{i\omega-P_i}\right]^{-1}
\end{equation}

For the left hand side
\begin{eqnarray}
  W_i &=\lim_{i\omega\rightarrow Q_i} \left(\sum^{N_S}_{j=1} \frac{W_j}{i\omega-Q_j}\right)(i\omega-Q_i) \nonumber \\
  &=\lim_{i\omega\rightarrow Q_i} \left[ i\omega+\mu-\epsilon_{imp}-\Sigma(\infty)-\left(\sum^{N_g}_{j=1}\frac{V_j}{i\omega-P_j}\right)^{-1}\right](i\omega-Q_i) \nonumber \\
  &=\lim_{i\omega\rightarrow Q_i} -(i\omega-Q_i)\left(\sum^{N_G}_{j=1}\frac{V_j}{i\omega-P_j}\right)^{-1}
\end{eqnarray}
here we define $\delta=i\omega-Q_i$ and $h(\delta)=\sum^{N_G}_{j=1}\frac{V_j}{Q_i+\delta-P_j}$, expand $h(\delta)$ as
\begin{equation}
  h(\delta) \approx \delta \cdot h'(0)=-\sum^{N_G}_{j=1}\frac{V_j}{(Q_i-P_j)^2}\cdot \delta
\end{equation}
so
\begin{equation}
  W_i=\left[ \sum^{N_G}_{j=1}\frac{V_j}{(Q_i-P_j)^2}\right]^{-1}
\end{equation}
here we have got both self energy poles and its weight.

\ack
The authors thank L.Wang, G. Xu, and F. Lu for their helpful discussions. We acknowledge the support 
from the 973 Program of China (Grant No. 2011CBA00108), from the NSF of China (Grants No. NSFC 10876042 and No. NSFC 10874158), and that from the Fund of Laboratory of Shock Wave and Detonation Physics, Institute of Fluid Physics, CAEP, Contract No. 2011-056000-0833F.

\section*{References}\label{references}

\bibliographystyle{unsrt}
\bibliography{papers}

\end{document}